\documentclass[%
 reprint,
 amsmath,amssymb,
 aps,
]{revtex4-2}

\usepackage{graphicx}
\usepackage{dcolumn}
\usepackage{bm}
\usepackage{upgreek}
\usepackage{amsmath}
\usepackage{mathrsfs}
\usepackage{mathtools}
\usepackage{xcolor}
\usepackage{cases}
\usepackage{ifthen}
\usepackage{soul}

\newcommand{\fpm}{f_{\mathrm{PM}}}

\newcommand{\abs}[1]{\left|#1\right|}
\newcommand{\comillas}[1]{``#1''}

\newcommand{\deff}{d_{\mathrm{eff}}}
\newcommand{\lcr}{\ell}
\newcommand{\lcav}{L_{\mathrm{cav}}}
\newcommand{\dtil}{\tilde{\delta}}

\newcommand{\derparvar}[2]{\frac{\partial #1}{\partial #2}}

\newcommand{\derparvarnil}[3]{\partial^{#1} #2/\partial #3^{#1}}
\newcommand{\trt}{t_\mathrm{rt}}
\newcommand{\dk}{\Delta k}
\newcommand{\sinc}[1]{\mathrm{sinc}\left( #1 \right)}

\newcommand{\fourier}[1]{\mathcal{F}\left\lbrace #1 \right\rbrace}
\newcommand{\invfourier}[1]{\mathcal{F}^{-1}\left\lbrace #1 \right\rbrace}

\newcommand{\Bin}{B_{\mathrm{in}}}

\newcommand{\Iin}{I_{\mathrm{in}}}
\newcommand{\Ith}{I_{\mathrm{th}}}

\newcommand{\fsr}{\Delta\nu}
\newcommand{\Lat}{\hat{L}_A^{\tau}}
\newcommand{\Lbt}{\hat{L}_B^{\tau}}
\newcommand{\Law}{\hat{L}_A^{\Omega}}
\newcommand{\Lbw}{\hat{L}_B^{\Omega}}
\newcommand{\ka}{\kappa_A}
\newcommand{\kb}{\kappa_B}
\newcommand{\Nreq}{N^{\mathrm{req}}}
\newcommand{\dcr}{\delta_{\mathrm{cr}}}
\newcommand{\phigvd}{\phi_{\mathrm{D}}}

\definecolor{mynaranja}{rgb}{0.84,0.32,0.10}
\definecolor{myrojo2}{rgb}{0.71,0 ,0}
\newboolean{EDITEDCOLOR}
\setboolean{EDITEDCOLOR}{false} 
\newcommand{\edited}[1]{%
    \ifthenelse{\boolean{EDITEDCOLOR}}%
        {\textbf{\textcolor{myrojo2}{#1}}}%
        {\textcolor{black}{#1}}%
}

\begin{document}

\preprint{APS/123-QED}

\title{Mean-field equation for phase-modulated optical parametric oscillator}

\author{A. D. Sanchez}
\email{alfredo.sanchez@icfo.eu}
\homepage{https://github.com/alfredos84/cuLLE}
\affiliation{ICFO-Institut de Ciencies Fotoniques, Mediterranean Technology Park, 08860 Castelldefels, Barcelona, Spain.}

\author{S. Chaitanya Kumar}

\affiliation{Tata Institute of Fundamental Research Hyderabad, 36/P Gopanpally, Hyderabad 500046, Telangana, India.}

\author{M. Ebrahim-Zadeh}
\affiliation{ICFO-Institut de Ciencies Fotoniques, Mediterranean Technology Park, 08860 Castelldefels, Barcelona, Spain.}
\affiliation{Instituciò Catalana de Recerca i Estudis Avancats (ICREA), Passeig Lluis Companys 23, Barcelona 08010, Spain.}


\date{\today}

\begin{abstract}
The widely established techniques for the generation of ultrashort optical pulses rely on passive mode-locking of lasers, with the output pulse duration and emission spectrum determined by the intrinsic lifetime of laser transition in the gain medium. Due to the instantaneous nature of nonlinear gain, optical parametric oscillators (OPOs) are capable of generating optical radiation in all time scales from continuous-wave (cw) to ultrashort femtosecond regime, if driven by laser pump sources in the corresponding time domain. In the ultrashort time scale, operation of OPOs conventionally relies on mode-locked pump lasers, with the concomitant disadvantages of large footprint and high cost. At the same time, the lack of gain storage mandates the use of synchronous pumping, resulting in increased complexity. In this paper, we present the concept of phase-modulated OPO driven by cw pump laser. The approach overcomes the traditional drawbacks of ultrafast OPOs, enabling femtosecond pulse generation without the need for synchronous pumping, resulting in a simplified, compact and cost-effective architecture using cw input pump lasers. We derive a mean-field equation for a degenerate $\chi^{(2)}$ OPO driven by a cw laser with intracavity electro-optic modulator (EOM), and also including dispersion compensation. The new equation predicts the formation of stable femtosecond pulses ($<$200~fs), \edited{in both normal and anomalous dispersion regimes}, with a controllable repetition rate determined by the frequency of the EOM. The remarkable functionality of the proposed scheme paves the way for the development of a new class of widely tunable coherent femtosecond light sources in both bulk and integrated format based on $\chi^{(2)}$ OPOs using cw pump lasers.
\end{abstract}

\maketitle

\section{Introduction}
\label{sec:intro}

The invention of mode-locked ultrafast lasers and their subsequent technological development has been a cornerstone of photonics, leading to the emergence of a vast multitude of applications from optical communications~\cite{TelecomApp} to medical diagnostics~\cite{MedicalApp}. Since the first demonstration of active mode-locking by synchronous intracavity modulation~\cite{FirstActiveModelocking}, and subsequently passive mode-locking using saturable absorbers~\cite{FirstPassiveSAModelocking} and Kerr-lens mode-locking~\cite{FirstKLM}, several mode-locking techniques have been investigated, paving the way for the successful realization of high-power ultrafast femtosecond lasers~\cite{ML_Review}. 

Despite the tremendous capability of mode-locking techniques to generate femtosecond pulses, ultrafast lasers typically operate at discrete wavelengths, or at best over a limited tuning range. This limitation imposed by the laser gain bandwidth can be overcome by $\chi^{(2)}$ nonlinear optical techniques based on bulk optical parametric oscillators (OPOs)~\cite{MEScience}, enabling expansive spectral coverage from the ultraviolet to mid-infrared~\cite{ebrahimzadeh2001optical, ebrahim2014yb2, ebrahim2014yb3}.  On the other hand, in the ultrashort picosecond and femtosecond time-scale, the instantaneous nature of the nonlinear gain in an OPO mandates the use of mode-locked ultrafast pump sources in combination with synchronous pumping~\cite{ebrahimzadeh2001optical}. Hence, the deployment of a mode-locked pump lasers has been a fundamental prerequisite for the development of widely tunable ultrafast OPOs, resulting in elaborate system design, large size, and high cost. A crucial step towards reducing the complexity, size, and cost, and harnessing the enormous potential of $\chi^{(2)}$ nonlinear parametric sources, would be to generate ultrashort femtosecond pulses directly from bulk continuous-wave (cw)-driven OPOs. Such an approach has been extensively investigated in dispersion-engineered microresonators exploiting $\chi^{(3)}$-based Kerr nonlinearity, where the generated signal and idler frequency combs in the four-wave mixing process lie symmetrically on either side and in close proximity to the cw pump frequency, generating femtosecond soliton pulses~\cite{Solitoncomb}. However, in a $\chi^{(2)}$ three-wave interaction within a quadratic medium, owing to the phase-matching properties of the nonlinear material, the driving pump frequency is significantly far from the generated signal/idler frequencies, making group velocity mismatch (GVM) and group velocity dispersion (GVD) management in these systems not only critical, but also indispensable for ultrashort pulse generation~\cite{nie2020quadraticDegOPO, nie2020quadraticSRO}. Although GVD compensation using intracavity prism pair and/or chirped mirrors in synchronously-pumped femtosecond OPOs is well-established, GVD control in cw OPOs is not common practice. Further, compared to Kerr nonlinearity, ultrashort pulse generation using quadratic nonlinearity also benefits from the intrinsically higher efficiency. To this extent, several approaches have been previously explored to generate ultrashort pulses from bulk cw OPOs~\cite{diddams1999broadband,forget2006actively,melkonian2007active,esteban2012frequency}. Using an intracavity acousto-optic modulator in a cw OPO resulted in the generation of 1~ns pulses at 1064~nm~\cite{forget2006actively,melkonian2007active}. On the other hand, by deploying an electro-optic phase modulator (EOM) internal to a green-pumped cw OPO, we demonstrated the generation of 230~ps pulses at 1064~nm~\cite{devi2013directly}. Further, quadratic frequency comb generation induced by modulation instability in a cw OPO was also demonstrated~\cite{mosca2018modulation}. \edited{The use of EOM in ring-resonators based on second- and third-order susceptibility has been very recently used to control the formation of dissipative solitons, higher order solitons or chaotic states~\cite{tusnin2020nonlinear, sun2023dynamics,englebert2023bloch}.} Recently, we showed that picosecond pulses down to $\sim$3.5 ps can be generated from a bulk cw OPO in the presence of an intracavity EOM~\cite{sanchez2022ultrashort}. However, the effect of dispersion compensation and its influence on the generation of femtosecond pulses was not explored in that work and, to the best of our knowledge, has not been previously investigated in the context of a bulk $\chi^{(2)}$ cw OPO. Very recently, a $\chi^{(2)}$ electro-optic comb generator in thin-film lithium niobate waveguide was demonstrated, generating 500-fs pulses at 30-GHz repetition rate by using an on-chip chirped Bragg grating to compensate for dispersion, but this was an integrated device based on single-pass interaction and in the absence of a resonant cavity~\cite{yu2022integrated}.

In this paper, we develop a theoretical framework for ultrashort pulse generation in $\chi^{(2)}$ cw-driven OPOs. We derive a mean-field equation (MFE) for such a device with intracavity EOM, also including dispersion compensation, as an extension of those derived in Refs.~\cite{lodahl1999pattern, mosca2018modulation}, and applicable to both bulk and integrated resonators. We focus our attention on a degenerate doubly-resonant OPO, since this configuration delivers the broadest output spectrum which is advantageous for numerous applications such as broadband spectroscopy and comb generation, in addition to enabling the attainment of shortest optical pulses. Our results show that this new MFE predicts the generation of ultrashort pulses ($<$200~fs) in both normal and anomalous dispersion regimes, as well as for zero temporal walk-off. Moreover, this single MFE provides complete control on the performance characteristics of the cw-driven degenerate OPO \edited{by accounting not only for the static detuning accumulated from the physical length of the cavity}, but also for the dynamic detuning resulting from phase-modulation provided by the EOM. As a result, the generated ultrashort pulses are accompanied by a variable repetition rate controlled by the modulation frequency of the EOM. 
The paper is organized as follows. In Section~\ref{sec:deriv}, we derive the MFE~\cite{lodahl1999pattern, zhou2014soliton, dong2021chirped}. In Section~\ref{sec:thresholdphase}, we focus on the threshold conditions and the phase properties of the signal. In Section~\ref{sec:EOM}, we detail some aspects related to the incorporation of the EOM and how it affects the threshold condition as well as the degeneracy of the OPO. In Section~\ref{sec:dispcomp}, we describe the formation of ultrashort pulses in different dispersion regimes. Finally, the conclusions of the work are presented in Section~\ref{sec:concl}.


\section{Derivation of the mean-field equation}
\label{sec:deriv}

The starting point in the development of a MFE for the signal electric field is the coupled-wave equations (CWEs) that well describe three-field interactions in nonlinear media. Since in degenerate parametric-down conversion (PDC) process both signal and idler are indistinguishable, the CWEs can be reduced to two equations. Let $A(z,\tau)$ be the resonant intracavity signal field at frequency $\omega_0$ and $B(z,\tau)$ the single-pass pump field at frequency $2\omega_0$ propagating through the nonlinear crystal, with the input external pump field amplitude denoted by $\Bin$. For a single-pass including linear absorption, group-velocity mismatch (GVM) and group-velocity dispersion (GVD), the CWEs read as
\begin{subequations}
\begin{eqnarray}
    \derparvar{A_m}{z} &= \Lat A_m + i\kappa_A B_m A_m^*e^{-i\Delta k z} \label{eq:CEs}\\
    \derparvar{B_m}{z} &=\Lbt B_m + i\kappa_B A_m^2e^{+i\Delta k z}, \label{eq:CEp}
\end{eqnarray}
\end{subequations}
where $\tau\in\left[-\trt/2,\trt/2\right]$ describes the temporal window during the round-trip time, $\trt$, in a co-moving frame with a group velocity, $\nu_{gA}$. The subscript $m$ labels the round-trip number, $z$ is spatial propagation coordinate along the nonlinear medium, and $\dk = 2k(\omega_0)-k(2\omega_0)$ is the phase-mismatch factor. The nonlinear coupling constant is  $\kappa_i=2\pi\deff/n_i\lambda_i$ ($i=A,~B$), with $\deff$ and $n_i$ being the effective nonlinear coefficient and refractive index of the nonlinear medium, respectively, evaluated at the corresponding wavelengths using the relevant Sellmeier equations. $\Lat$ and $\Lbt$ are the corresponding linear operators expressed in time ($\tau$) and frequency ($\Omega$) domain, given by
\begin{widetext}
\begin{subequations}
\begin{gather}
    \Lat = -\left( \frac{\alpha_{cA}}{2} +i\frac{k^{''}_A}{2}\frac{\partial^2}{\partial \tau^2} \right) \xleftrightarrow[]{\mathcal{F}} \Law = -\left( \frac{\alpha_{cA}}{2} -i\frac{k^{''}_A}{2}\Omega^2 \right) \label{eq:linops} \\
    \Lbt = -\left( \frac{\alpha_{cB}}{2}+\Delta k^{\prime}\frac{\partial}{\partial \tau}+i\frac{k^{''}_B}{2}\frac{\partial^2}{\partial \tau^2} \right) \xleftrightarrow[]{\mathcal{F}} \Lbw = -\left( \frac{\alpha_{cB}}{2}-i\Delta k^{\prime}\Omega-i\frac{k^{''}_B}{2}\Omega^2 \right) \label{eq:linopp},
\end{gather}
\end{subequations}
\end{widetext}
where $\fourier{\cdot}=\int_{-\infty}^{\infty}\cdot e^{i\Omega}d\tau$ stands for the Fourier transform and the transformation $\partial/\partial\tau\leftrightarrow -i\Omega$ has been used. Here, $\alpha_{ci}$ is the linear absorption, $\Delta k^{\prime} = \nu^{-1}_{gB} -\nu^{-1}_{gA}$ is the GVM, and $k^{''}_i=\left.\derparvarnil{2}{k}{\omega}\right|_{\omega_i}$ is the GVD. The CWEs must be solved over the entire nonlinear medium of length, $\lcr$.

\begin{figure}[hbpt]
    \centering
    \includegraphics[width=1.0\linewidth]{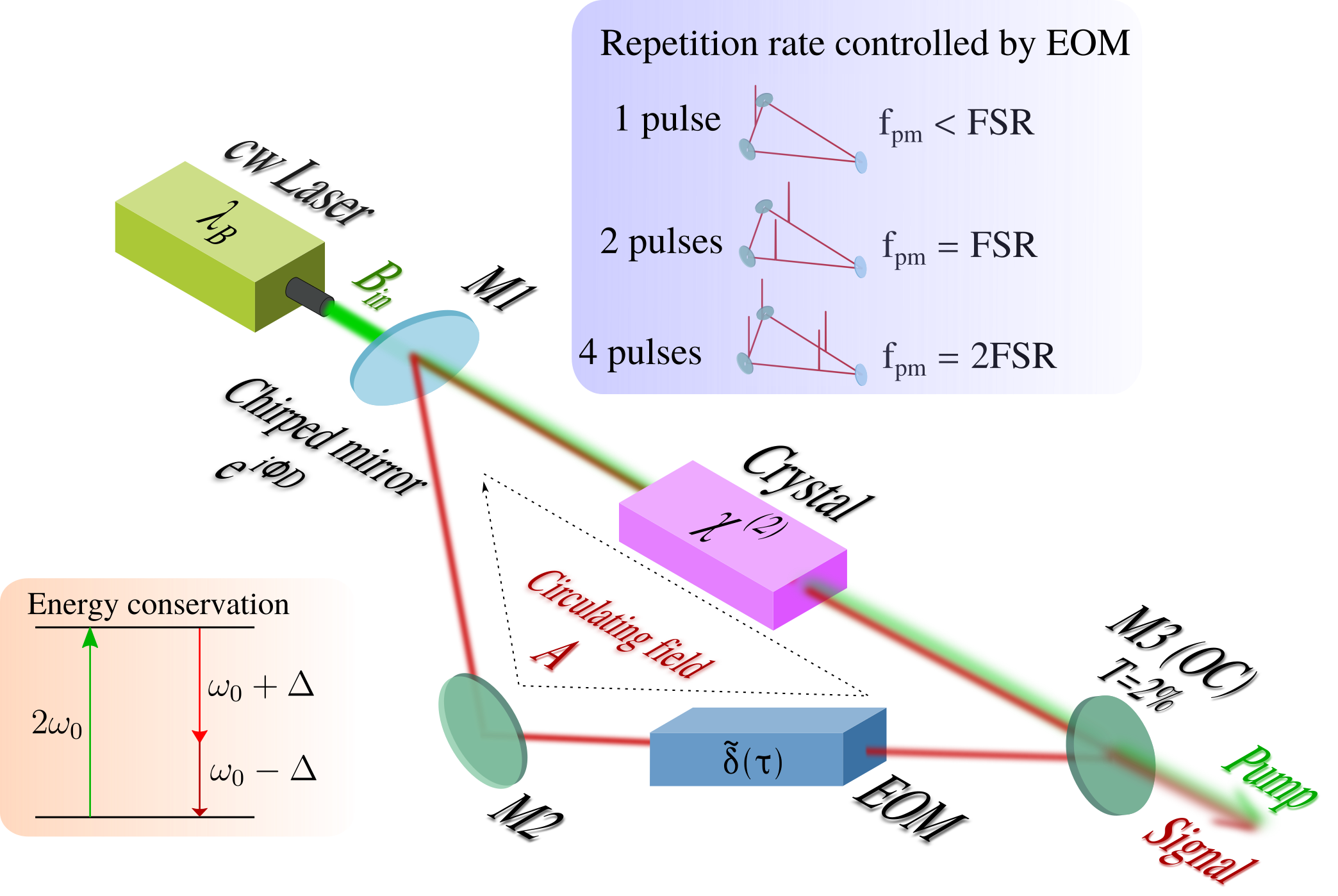}
    \caption{Schematic of the bulk OPO used in our model. The mirrors M1, M2 and M3 form a ring cavity where the nonlinear crystal is placed with an intracavity electro-optic modulator (EOM). The mirror M1 acts as an input coupler, while the chirped mirror, M3, compensates for the group-velocity dispersion ($e^{i\phigvd}$), and is the output coupler with a transmission of 2\%. (left inset) Energy conservation diagram. Here, $\Delta\ll\omega_0$ represents a small deviation from degeneracy, enabled by the cavity detuning. (right inset) The frequency of the EOM, $\fpm$, controls the amount of intracavity pulses (FSR is the cavity free-spectral-range), and also acts as a time-\textcolor{black}{dependent} cavity detuner, $\dtil(\tau)$.}
    \label{fig:scheme}
\end{figure}

After one round-trip, the electric fields must be updated depending on cavity losses and all intracavity elements. After reflection from the output coupler, the electric fields attenuate as $A_m(z^{\mathrm{OC}},\tau)\rightarrow \sqrt{R}A_m(z^{\mathrm{OC}},\tau)$, with $z^{\mathrm{OC}}$ the output coupler position, $R=1-\theta_A$ the power reflectivity, and $\theta_A$ the power transmittance at $\omega_0$. To incorporate the EOM as well as the cavity detuning, the proper phase should be added as $\sqrt{R}A_m(z^{\mathrm{EOM}},\tau) \rightarrow \sqrt{R}e^{i\dtil(\tau)}A_m(z^{\mathrm{EOM}},\tau)$, with $z^{\mathrm{EOM}}$ the EOM position in the cavity, where 
\begin{equation}\label{eq:tdet}
\dtil(\tau)= \beta\sin(2\pi\fpm\tau)-\delta    
\end{equation}
is the net time-dependant cavity detuning, $\delta=(\omega_0-\omega_{\mathrm{cav}})\trt$, with $\omega_{\mathrm{cav}}$ the cavity mode frequency, and $\beta$ and $\fpm$ the modulation depth and the frequency modulation of the EOM, respectively. Finally, we are interested in the GVD compensation before starting the next round-trip. This is done by adding to the electric field a quadratic phase in the frequency domain as $\fourier{\sqrt{R}e^{\dtil(\tau)}A_m(\lcav,\tau)} \rightarrow \fourier{\sqrt{R}e^{\dtil(\tau)}A_m(\lcav,\tau)}e^{i\phigvd}$~\cite{agrawal2000nonlinear}. Thus, using the convolution theorem, the updated fields at the beginning of the $(m+1)-$round-trip and the intracavity fields are related to the $m-$round-trip as
\begin{subequations}
\begin{eqnarray}
A_{m+1}(0,\tau) &= \left[\sqrt{1-\theta_A}e^{i\dtil(\tau)}A_m(\lcav,\tau)\right]\otimes\tilde{\Phi}_{\mathrm{D}},\label{eq:BCs}  \\
B_{m+1}(0,\tau) &= \Bin,\label{eq:BCp}
\end{eqnarray}
\end{subequations}
where $\theta_A=1-R$ is the power transmittance, $\lcr$ is the crystal length, and $\phigvd = \gamma\Omega^2/2$. The parameter, $\gamma=-\epsilon k^{''}_A\lcr$, takes into account the GVD compensation at the signal frequency, $\epsilon\in \left[0,1\right]$ is the compensation index, and $\tilde{\Phi}_{\mathrm{D}} = \invfourier{e^{i\phigvd}}$ . Notice that we have arranged all intracavity elements in $z=\lcav$, because we are working in the plane-wave approximation. Also, the mathematical formulations for these elements are linear operations and can commute with one another, independent of their position in the cavity. Figure~\ref{fig:scheme} shows the physical system used for our modelling: a cw-driven bulk $\chi^{(2)}$ OPO including intracavity EOM and dispersion compensation.

In order to derive the single MFE, we firstly solve the Eq.~\ref{eq:CEp} following the steps in Refs.~\cite{haelterman1992dissipative, valiulis2011propagation, zhou2014soliton}. The goal is to combine Eqs.~\ref{eq:CEs}-\ref{eq:CEp}-\ref{eq:BCs}-\ref{eq:BCp}, often called \comillas{infinite-dimensional map}~\cite{hansson2017singly}, into a single equation that describes the evolution of the signal electric field driven by an external field, $\Bin$. We make the substitution, $B=Ce^{+i\dk z}$, in Eq.~\ref{eq:CEp} (we will omit the subscript $m$, since the electric field $B$ does not resonate in the cavity), resulting in
\begin{equation}
    \frac{\partial}{\partial z}\left(Ce^{+i\dk z}\right) = \Lbt Ce^{+i\dk z} + i\kb A^2e^{+i\Delta k z}
\end{equation}
\begin{equation}
    \left(\derparvar{C}{z}+i\dk C\right) e^{+i\dk z} =\left(\Lbt C + i\kb A^2\right) e^{+i\Delta k z}.
\end{equation}
By canceling the factors, $e^{+i\dk z}$, and taking the Fourier transform, we obtain
\begin{equation}\label{eq:eqc}
    \derparvar{\tilde{C}}{z} + D(\Omega) \tilde{C} = i\kb \fourier{A^2},
\end{equation}
with 
\begin{equation}
    D(\Omega) =i\dk - \Lbw.
\end{equation}
The most general solution for Eq.~\ref{eq:eqc} is 
\begin{equation}\label{eq:eqcgeneral}
    \tilde{C} = \tilde{C}^{\mathrm{H}} + \tilde{C}^{\mathrm{P}},
\end{equation}
where the superscripts H and P stand for homogeneous and particular solutions, respectively.

Homogeneous solution: To give the differential equation homogeneity, we have to set the driving term, $i\kb\fourier{A^2}$, to zero,
\begin{equation}
    \frac{\partial \tilde{C}^{\mathrm{H}}}{\partial z} + D(\Omega) \tilde{C}^{\mathrm{H}} = 0 \Rightarrow \tilde{C}^{\mathrm{H}} = \tilde{C}^{\mathrm{H}}_0 e^{-D(\Omega)z},
\end{equation}
where $\tilde{C}^{\mathrm{H}}_0$ is given by the initial condition.

Particular solution: we consider a solution independent of $z$, so that $\partial \tilde{C}^{\mathrm{P}}/\partial z = 0$. In this case, we obtain
\begin{equation}
    D(\Omega) \tilde{C}^{\mathrm{P}} = i\kb \fourier{A^2} \Rightarrow  \tilde{C}^{\mathrm{P}} = \frac{i\kb \fourier{A^2}}{D(\Omega)}.
\end{equation}

Boundary condition: assuming that at $z=0$,  $\tilde{C}(0,\Omega)=\Bin\delta(\Omega)$, with $\delta$ the Dirac function, from Eq.~\ref{eq:eqcgeneral} we obtain
\begin{equation}
    \tilde{C}(0, \Omega) = \tilde{C}^{\mathrm{H}}_0 e^{-D(\Omega)0}  + \frac{i\kb \fourier{A^2}}{D(\Omega)} =\Bin\delta(\Omega),
\end{equation}
thus
\begin{equation}
    \tilde{C}^{\mathrm{H}}_0 = \Bin\delta(\Omega) -\frac{i\kb \fourier{A^2}}{D(\Omega)}.
\end{equation}
By combining these results, we obtain
\begin{equation}
    \tilde{C}(z,\Omega) = \Bin\delta(\Omega)e^{-D(\Omega)z} +  \frac{i\kb\fourier{A^2}}{D(\Omega)}\left(1-e^{-D(\Omega)z}\right),
\end{equation}
and multiplying by $e^{+i\dk z}$ to recover $\tilde{B}(z,\Omega)=\tilde{C}(z,\Omega)e^{+i\dk z}$, we arrive at
\begin{equation}
    \tilde{B} = \Bin e^{\Lbw z}\delta(\Omega) + \frac{i\kb\fourier{A^2}}{i\dk - \Lbw}\left(e^{+i\dk z} - e^{\Lbw z} \right).
\end{equation}
By assuming the pump losses in the nonlinear medium are negligible ($\alpha_{cB}=0$) and taking the inverse Fourier transform of the previous expression, we obtain
\begin{equation}\label{eq:Bfield}
    B \approx \Bin + \invfourier{i\kb \fourier{A^2}\frac{e^{i\dk z}- e^{\Lbw z}}{i\dk-\Lbw}}.
\end{equation}
The final step is to introduce Eq.~\ref{eq:Bfield} into Eq.~\ref{eq:CEs}, and then integrate over the full interaction length
\begin{widetext}
\begin{equation}\label{eq:intArt}
\int_{0}^{\lcr}\derparvar{A}{z} \,dz =\int_{0}^{\lcr} \Lat A + i\ka \invfourier{i\kb \fourier{A^2}\frac{e^{i\dk z}- e^{\Lbw z}}{i\dk-\Lbw}}A^* e^{-i\Delta k z} + i\ka\Bin A^* e^{-i\Delta k z} \,dz.
\end{equation}
\end{widetext}
The left-side in Eq.~\ref{eq:intArt} can be approximated as
\begin{equation}\label{eq:intArt1}
    \int_{0}^{\lcr}\derparvar{A}{z} \,dz \approx A(\lcr,\tau) - A(0,\tau), 
\end{equation}
whilst in the terms on the right-side we assume $A(z,\tau)$ and $A^*(z,\tau)$ \edited{to be constant in the integration.} The first term on the right-side is 
\begin{equation}\label{eq:intArt2}
    \int_{0}^{\lcr} \Lat A \,dz =  \lcr\Lat A.
\end{equation}
In the second term we also assume $\fourier{A^2(z,\tau)}$ to be constant in the integration. Using $\Gamma=\dk+i\Lbw$, we obtain
\begin{widetext}
\begin{equation}\label{eq:intArt3}
    \int_{0}^{\lcr} i\ka \invfourier{i\kb \fourier{A^2}\frac{e^{i\dk z}- e^{\Lbw z}}{i\dk-\Lbw}}A^* e^{-i\Delta k z}\,dz = i\ka\kb A^*\invfourier{i\fourier{A^2}\int_{0}^{\lcr}\frac{1- e^{-i\Gamma z}}{i\Gamma}\,dz},
\end{equation}    
\end{widetext}
The integral in Eq.~\ref{eq:intArt3} is
\begin{equation}\label{eq:intArt4}
\int_{0}^{\lcr} \frac{1- e^{-i\Gamma z}}{\Gamma}\,dz = \lcr^2\frac{1-ix(\Omega)-e^{-ix(\Omega)}}{-ix(\Omega)^2},
\end{equation}
with $x(\Omega)=\Gamma \lcr$. Therefore, Eq.~\ref{eq:intArt3} can be written as
\begin{equation}\label{eq:intArt5}
    -\rho^2  A^*(z,\tau) \left[ A^2(z,\tau)\otimes I(\tau)\right],
\end{equation}
with $\rho^2=\ka\kb\lcr^2$. Here, the convolution theorem has been used and the function $I(\tau)=\invfourier{\hat{I}(\Omega)}$, with
\begin{equation}\label{eq:intArt6}
    \hat{I}(\Omega) = \frac{1-ix(\Omega)-e^{-ix(\Omega)}}{x(\Omega)^2},
\end{equation}
is known as the delayed response and resembles the Raman response in Kerr media~\cite{stolen1989raman, hansson2017singly}. 

Finally, the third term is
\begin{equation}\label{eq:intArt7}
 \int_{0}^{\lcr} i\ka A^*(z,\tau) \Bin e^{-i\dk z}\,dz = i\sigma(\eta)A^*(0,\tau),
\end{equation}
where $\sigma(\eta)=\ka\lcr\Bin g(\eta)$, $g(\eta)=e^{-i\eta}\sinc{\eta}$, and $\eta=\dk\lcr/2$. 
By combining Eqs.~\ref{eq:intArt1}-\ref{eq:intArt7}, we obtain
\begin{widetext}
\begin{equation}\label{eq:intArt8}
    A(\lcr,\tau) = A(0,\tau) +  \lcr\Lat A(0,\tau) -\rho^2  A^*(0,\tau) \left( A^2(0,\tau)\otimes I(\tau)\right) + i\sigma(\eta) A^*(0,\tau). 
\end{equation}
\end{widetext}
Finally, by substituting Eq.~\ref{eq:intArt8} in Eq.~\ref{eq:BCs} and assuming $\theta_A$ and $\dtil(\tau)$ as first-order quantities, with
\begin{equation}
    \sqrt{1-\theta_A}e^{i\dtil(\tau)}\approx 1 - \frac{\theta_A}{2} + i\dtil(\tau),
\end{equation}
obtaining
\begin{widetext}
\begin{multline}
A_{m+1}(0,\tau)-A_{m}(0,\tau)\otimes\tilde{\Phi}_{\mathrm{D}}=\\
    \left\lbrace\left(i\dtil(\tau)-\alpha_A-i\frac{k^{''}_A\lcr}{2}\frac{\partial^2}{\partial \tau^2} \right)A_m(0,\tau)
    -\left( \rho^2 \left( A^2_m(0,\tau)\otimes I(\tau)\right) - i\sigma(\eta) \right)A_m^*(0,\tau)\right\rbrace \otimes\tilde{\Phi}_{\mathrm{D}},
\end{multline}
\end{widetext}
where $\alpha_A = (\theta_A+\alpha_{cA}\lcr)/2$. Finally, we obtain the single MFE for the signal electric field, $A=A(t,\tau)$, by defining the slow-time derivative as
\begin{equation}
    \trt\derparvar{A}{t} \coloneqq A_{m+1}(0,\tau)-A_{m}(0,\tau)\otimes\tilde{\Phi}_{\mathrm{D}},
\end{equation}
and hence
\begin{widetext}
\begin{equation}\label{eq:LLE_GDD_EOM}
    \trt\derparvar{A}{t} = \left\lbrace \left(i\dtil(\tau)-\alpha_A-i\frac{k^{''}_A\lcr}{2}\frac{\partial^2}{\partial \tau^2} \right)A-\left( \rho^2 \left( A^2\otimes I(\tau)\right) - i\sigma(\eta) \right)A^* \right\rbrace \otimes\tilde{\Phi}_{\mathrm{D}}.
\end{equation}
\end{widetext}
As expected, in the absence of intracavity EOM and dispersion compensation ($\dtil(\tau)=\delta$ and $\epsilon=0$, respectively), Eq.~\ref{eq:LLE_GDD_EOM} becomes the standard MFE derived in previous works ~\cite{mosca2018modulation}. An analytical solution for Eq.~\ref{eq:LLE_GDD_EOM} is not available due to its complexity, and numerical methods are indispensable for its analysis. \edited{It is worth mentioning that the consideration of third-order dispersion (TOD) or higher is necessary as long as a prior evaluation justifies its inclusion. In the experimental scheme considered in this work, we have confirmed that the role of TOD is negligible and inclusion of this parameter is not necessary in the model.} Also notice that, strictly speaking, we should have added an extra factor, $\delta(\tau-\trt/2)$, in Eq.~\ref{eq:LLE_GDD_EOM}, to specify that the convolution is only computed at the end of the round-trip time (the symbol, $\delta$, here stands for Dirac function and not for cavity detuning). We numerically solve this equation by using a standard four-order Runge-Kutta method. The convolution terms are solved using the convolution theorem, while the dispersion term is solved in frequency domain. In our numerical calculations, we use data for lithium niobate nonlinear crystal available in Ref.~\cite{gayer2008temperature} as the nonlinear medium of length, $\lcr=5$~mm.

\section{Threshold condition and phase properties}
\label{sec:thresholdphase}

The analysis developed in Ref.~\cite{mosca2018modulation} does not include an EOM or dispersion compensation ($\dtil(\tau)=\delta$ and $\gamma=0$), and the MFE
\begin{widetext}
\begin{equation}\label{eq:LLE_RAW}
    \trt\derparvar{A}{t} = \left(i\delta-\alpha_A-i\frac{k^{''}_A\lcr}{2}\frac{\partial^2}{\partial \tau^2} \right) A-\left( \rho^2 \left( A^2\otimes I(\tau)\right) - i\sigma(\eta) \right)A^* 
\end{equation}
\end{widetext}
yields the threshold condition and the phase of the signal electric field. For such purpose, an anzatz $A(t,\tau)=\abs{A_0}e^{i\phi_A}$ is proposed as a solution of Eq.~\ref{eq:LLE_RAW} in the steady-state ($\partial A/\partial t=0$), obtaining
\begin{equation}\label{eq:A02}
    \abs{A_0}^2 = \frac{-\alpha_A\pm\sqrt{(\ka\lcr\Bin)^2-\delta^2}}{\ka\kb\lcr^2\hat{I}(0)},
\end{equation}
and
\begin{equation}\label{eq:cosphia}
    \cos{2\phi_A} = \frac{\delta}{\ka\lcr\Bin}.
\end{equation}
Equation~\ref{eq:A02} yields to the standard threshold condition
\begin{equation}\label{eq:bin}
    \Bin^2 \geq \frac{\alpha_A^2+\delta^2}{\ka^2\lcr^2},
\end{equation}
where we assume perfect phase matching, $\eta=0$ ($g(0)=1$). The threshold intensity for a non-detuned ($\delta=0$) and doubly-resonant cavity is given by~\cite{ebrahimzadeh2001optical}
\begin{equation}\label{eq:Ith}  
    \Ith = \frac{\epsilon_0cn_Bn_A^2\lambda_A^2 }{8\pi^2\deff^2\lcr^2}\alpha_A^2.
\end{equation} 
It is convenient to use the pumping level defined as
\begin{equation}
    N = \frac{\Iin}{\Ith},
\end{equation}
where the input power is $\Iin = \epsilon_0cn_B\abs{\Bin}^2/2$. After some algebraic steps from Eq.~\ref{eq:bin}, we obtain
\begin{equation}\label{eq:Nreq}
    \Nreq\coloneqq\frac{\Iin}{\Ith} \geq 1+\left(\frac{\delta}{\alpha_A}\right)^2,
\end{equation}
where $\Nreq$ represents the required pumping level to switch on the OPO for a given cavity detuning. 

\begin{figure}[htbp]
    \centering
    \includegraphics[width=1.0\linewidth]{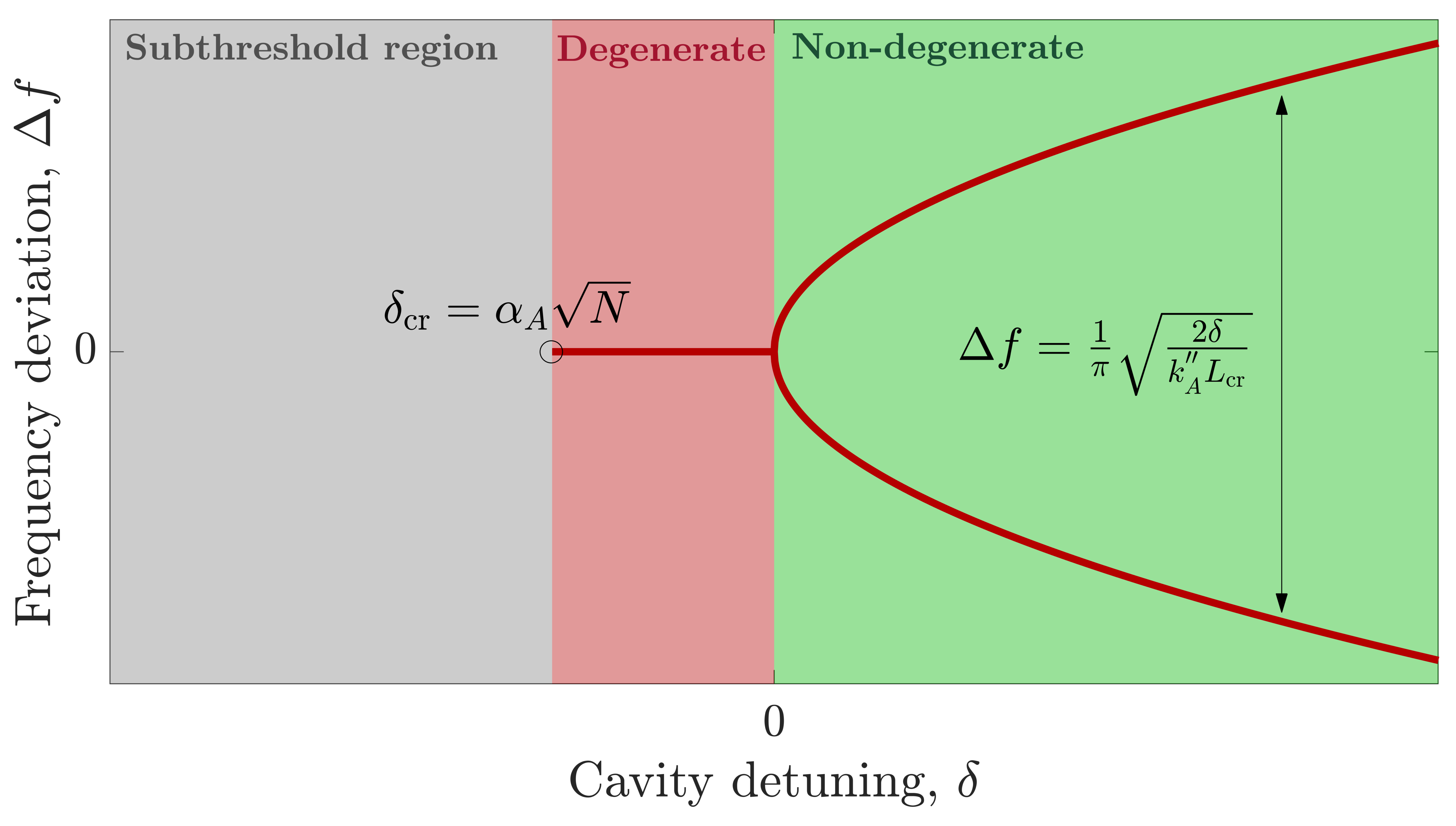}
    \caption{Spectral behaviour as a function of the cavity detuning: subthreshold, degenerate and non-degenerate OPO regimes. This sketches a regime where the signal fields operates in the normal dispersion regime ($k_A^{''}>0$). The fork-like spectral behaviour of the OPO flips for anomalous dispersion regime ($k_A^{''}<0$.)}
    \label{fig:fork}
\end{figure}

\begin{figure*}[ht]
    \centering
    \includegraphics[width=1.0\textwidth]{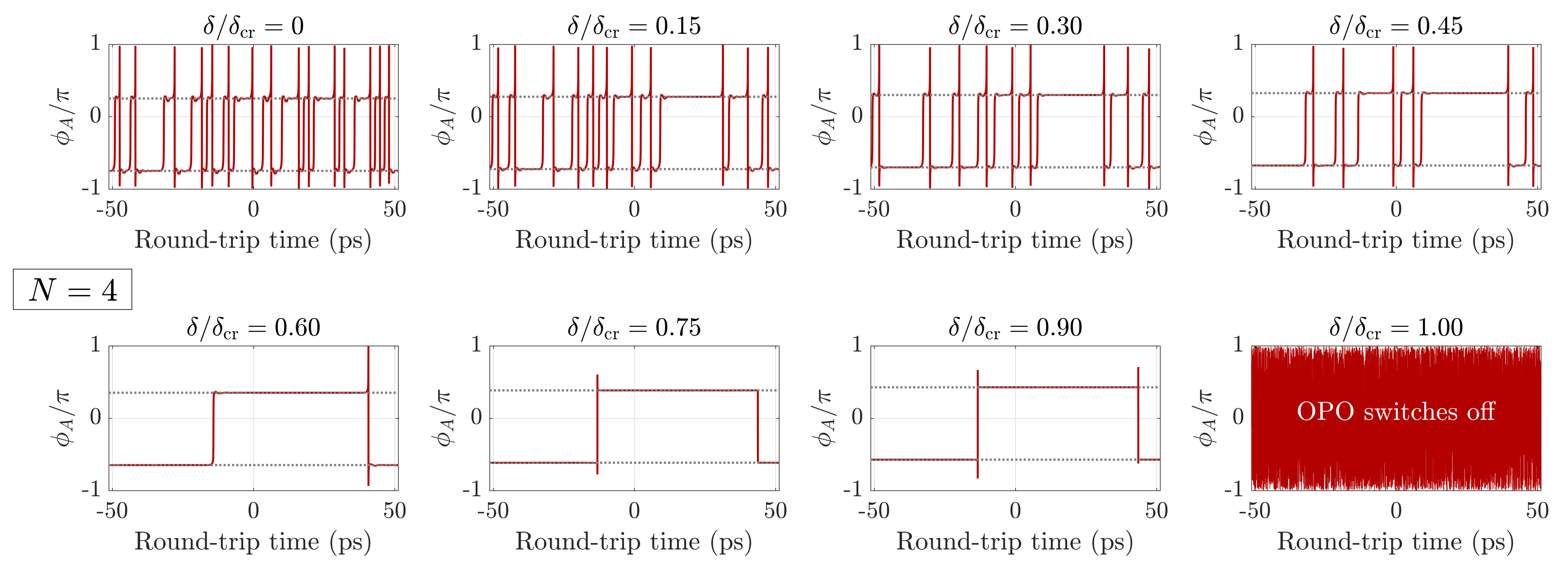}
    \caption{Signal phase as a function of the round-trip time for various cavity detuning in the degenerate regime. The horizontal lines correspond to the average values that the signal phase can take, given by Eq.~\ref{eq:phivsdelta}.}
    \label{fig:phasesN4}
\end{figure*}

A linear stability analysis on Eq.~\ref{eq:LLE_RAW} reveals that the OPO can oscillate despite the cavity being detuned, albeit less strongly due to the dependence of $N$ on $\delta$ (Eq.~\ref{eq:Nreq}). From the definition of modulation instability gain, it is found that the OPO operates in degenerate mode for $\delta\leq 0$, while for $\delta>0$ there is a frequency splitting whose frequency separation is~\cite{smith2003degenerate}
\begin{equation}\label{eq:dfseparation}
    \Delta f = \frac{1}{\pi}\sqrt{\frac{2\delta}{k^{''}_A\lcr}},
\end{equation}
and hence the signs of $\delta$ and $k^{''}_A$ will determine whether or not the OPO will operate in degenerate mode once the cavity is detuned. On the other hand, for negative detuning and from Eqs.~\ref{eq:cosphia} and~\ref{eq:Ith}, it is deduced that $\ka\lcr\Bin=\alpha_A\sqrt{N}$. Since $\phi_A$ is a real quantity, $\delta/\dcr \leq 1$ must be fulfilled, where $\dcr = \alpha_A\sqrt{N}$ is the critical detuning value from which the OPO, in degenerate mode, ceases operation and switches off. As a consequence, the phase of the degenerate electric field is
\begin{equation}\label{eq:phivsdelta}
    \phi_A = \frac{1}{2}\arccos\left(\frac{\delta}{\dcr}\right).
\end{equation}
Figure~\ref{fig:fork} schematically shows the regimes in which the OPO can operate. At the critical point of non-detuned cavity, $\delta=0$, the OPO exhibits a second-order phase transition in the spectral domain between degenerate and non-degenerate regimes~\cite{roy2021spectral}.
The OPO can sustain the degenerate state for $\delta\in\left[-\dcr,0\right]$, with $\dcr=\alpha_A\sqrt{N}$. This means the stability of the OPO in the degenerate regime can be improved by increasing the pumping level, $N$, or increasing (decreasing) the transmittance, $\theta$ (reflectivity, $R$). For an OPO whose cavity has $\delta=0$, the phase relation between the pump and the signal remains fixed according to~\cite{wong2008self}
\begin{equation} \label{eq:spl}
    \phi_B-2\phi_A+\frac{\pi}{2} = 2q\pi,
\end{equation}
where $q$ is an integer, and $\phi_A$ and $\phi_B$ are the pump and signal phases, respectively. By regarding $\phi_B=0$, the average value of the signal phase is always found to be $\pi/4$ or~$-3\pi/4$. However, according to Eq.~\ref{eq:phivsdelta}, the values for the signal phase will change as the cavity detunes, for a fixed pumping level $N$. Figure~\ref{fig:phasesN4} shows the temporal phase during the full round-trip time for different values of the normalized cavity detuning and for a fix pumping level, $N=4$. It is evident that as the normalized detuning, $\delta/\dcr\rightarrow 1$, the intermittent oscillations of the signal phase are increasingly slower. When $\delta/\dcr=1$, the OPO switches off and the phase signal assumes random values during propagation. The importance of the results in Fig.~\ref{fig:phasesN4} is that even when the cavity exhibits a non-zero detuning at degeneracy, the signal phase is locked \edited{between} two $\pi$-separated states. In the non-degenerate state, Eq.~\ref{eq:spl} is still valid, but with the replacement, $2\phi_A \rightarrow \phi_{\mathrm{signal}}+\phi_{\mathrm{idler}}$~\cite{ebrahimzadeh2001optical}.

\section{Electro-optic modulator}
\label{sec:EOM}

\begin{figure*}[htbp]
\centering
\includegraphics[width=1.0\textwidth]{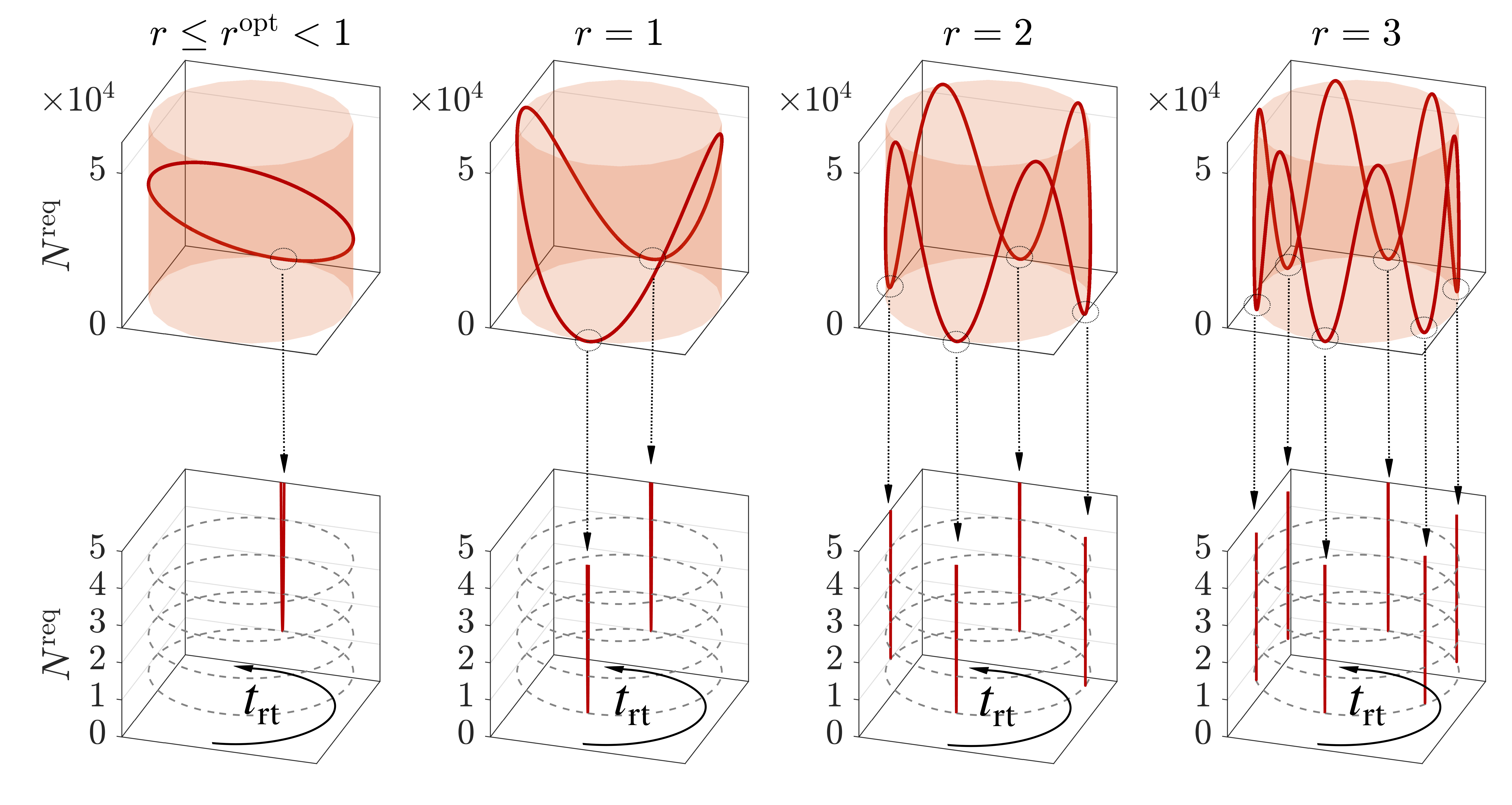}
    \caption{Required pumping level, $N^{\mathrm{req}}$, during the full round-trip for different $\fpm=r\fsr$. (top row) The required threshold for different values of $r$. (bottom row) A more detailed visualization of the required threshold for practical values of the pumping level, where the threshold occurs in a very short time window. The gray dashed circles represent pumping levels of $N =1, 2, 3, 4$. to highlight the fact that the threshold is only limited to a very small fraction of the total round-trip.}
    \label{fig:reqthreshold}
\end{figure*}

The EOM can be treated as a time-dependant detuning element, given its mathematical representation (Eq.~\ref{eq:tdet}). The deployment of EOM is often associated with a spectral broadening induced on the signal field~\cite{diddams1999broadband}. The spectral bandwidth is given by Eq.~\ref{eq:dfseparation}, but with the replacement, $\delta\leftrightarrow\beta$, while the maximum broadening is limited by the parametric gain bandwidth~\cite{sanchez2022ultrashort}. In other words, the EOM spans the fork shown in Fig.~\ref{fig:fork} in the range of detuning $\pm\beta$ along the entire round-trip time, from degenerate to non-degenerate state, expanding the spectrum. Most importantly, the inclusion of the EOM makes the required threshold in Eq.~\ref{eq:Nreq} time-dependent, that is
\begin{equation}\label{eq:Nreq_EOM}
    \Nreq(\tau) \geq 1+\left(\frac{\beta\sin\left(2\pi\fpm\tau\right)}{\alpha_A}\right)^2,
\end{equation}
where $\fpm$ is the modulation frequency, whose value is often close to FSR of the cavity, $\fpm=r\fsr$, with $r$ a positive number~\cite{esteban2012frequency, devi2013directly}. Note that the static detuning was neglected ($\delta=0$). The deployment of an EOM allows us to modify the threshold condition during the full round-trip. For a given set of control parameters $(N,\beta,r)$, Eq.~\ref{eq:Nreq_EOM} is clearly fulfilled only in specific intervals of $\tau$ over the full round-trip. To clarify this, Fig.~\ref{fig:reqthreshold} shows the right-side of the Eq.~\ref{eq:Nreq_EOM} for fixed values of $\beta=0.8\pi$ and $\alpha_A=0.01$, for various $r$. In the top row, we can see that $N^{\mathrm{req}}$ takes values from 1 to $\sim 10^4$.
For typical pumping values (e.g. $N\sim 1$), the threshold condition is fulfilled only during a very small fraction of the total round-trip time. This can be seen in the bottom row, where $N^{\mathrm{req}}\leq 5$ is plotted. Dashed circles represent pumping levels of $N =1, 2, 3, 4$. The time window in which the threshold condition is satisfied is on the order of a few femtoseconds for a round-trip of $\approx100$~ps, and increases slightly if the pumping level is increased and/or when the value of $r$ increases. 

Notice that to form a single time window for the threshold condition, it is necessary to set the modulation frequency in such a way that the threshold of the second window is prevented, as in the case for $r=1$. The calculation of the minimum factor required, $r^{\mathrm{opt}}$, can be derived from Eq.~\ref{eq:Nreq_EOM} by solving the inequality for $r$ and assessing $\tau=\trt/2$ (the end of the round-trip time window), yielding
\begin{equation}\label{eq:roptim}
    r^{\mathrm{opt}} = \frac{1}{\pi}\arcsin\left(\frac{\alpha_A}{\beta}\sqrt{N-1}\right)<1.
\end{equation}

\section{Dispersion compensation and pulse formation}
\label{sec:dispcomp}

Despite the fact that the EOM enables photon conversion from pump to signal over short-time windows, $\delta\tau^{\mathrm{w}}$, which can be computed from Eq.~\ref{eq:Nreq_EOM} as
\begin{equation}
    \delta\tau^{\mathrm{w}} \leq \abs{\frac{1}{2\pi\fpm}\arcsin\left(\frac{\alpha_A}{\beta}\sqrt{N-1}\right)},
\end{equation}
the presence of the cavity allows for multiple oscillating modes. Due to group velocity dispersion, different spectral components have different velocities. Therefore, total constructive interference between the majority of modes will not be possible and, therefore, phase alignment is required for eventual pulse formation. This is achieved by fully compensating for dispersion on each round-trip; that is, an extra phase, $\phigvd=-k^{''}_A\Omega^2\lcr/2$, is added in every round-trip. It is noteworthy that dispersion compensation does not imply that the entire nonlinear medium has zero GVD; a phase of opposite sign is added after a single pass in order to achieve a net zero GVD in the entire system. 

\begin{figure}[htbp]
    \centering
    \includegraphics[width=1.0\linewidth]{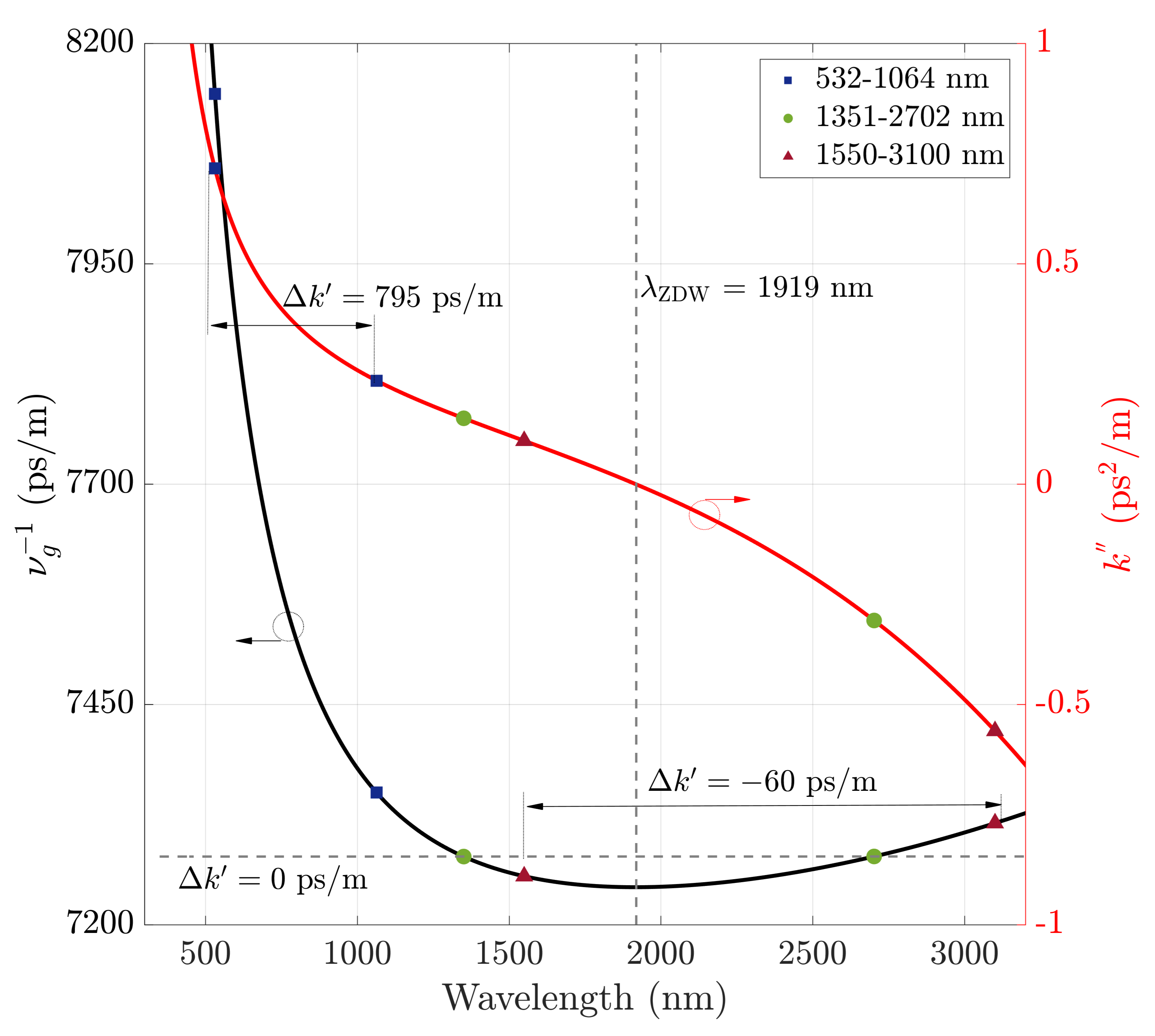}
    \caption{Dispersion properties of MgO:PPLN with the relevant pair of pump-signal wavelengths, indicating the dispersion regimes and the values for GVM.}
    \label{fig:disp_prop}
\end{figure}

Before presenting numerical solutions of the Eq.~\ref{eq:LLE_GDD_EOM} under the aforementioned conditions, it is noteworthy to point out the dispersion properties of the nonlinear medium used in our modeling. We have chosen wavelengths with the purpose of investigating the OPO behavior in the normal (532-1064~nm) and anomalous dispersion regime (1550-3100~nm), as well as the zero GVM condition (1351-2702~nm, $\dk^{\prime}=0$). This can be seen in Fig.~\ref{fig:disp_prop}, where the pump-signal pairs of wavelengths on the group velocity and GVD curves are shown schematically. The selected nonlinear medium corresponds to MgO-doped periodically-poled lithium niobate (MgO:PPLN)~\cite{gayer2008temperature}, which has a zero-dispersion wavelength $\lambda_{\mathrm{ZDW}}=1919$~nm (vertical dashed line). As can be seen, for a pump at 532~nm (1550~nm), the signal falls into the normal (anomalous) dispersion regime. For a pump at 1351~nm, the signal at 2702~nm has a group velocity identical to that of the pump (dashed horizontal line). At this particular wavelength, the system exhibits zero GVM. 

Based on the dispersion properties of the nonlinear medium, we find the numerical solutions of Eq.~\ref{eq:LLE_GDD_EOM}, for a fixed pumping level of $N=4$ and $\beta=0.8\pi$. The reason for this specific value is that at $N=4$ pumping level the highest conversion efficiency~\cite{yang1993power} is achieved. Simulations are truncated after $2\times 10^4$ round-trips in order to ensure that the system has reached the steady-state. 

\begin{figure*}[htbp]
    \centering
    \includegraphics[width=1.0\textwidth]{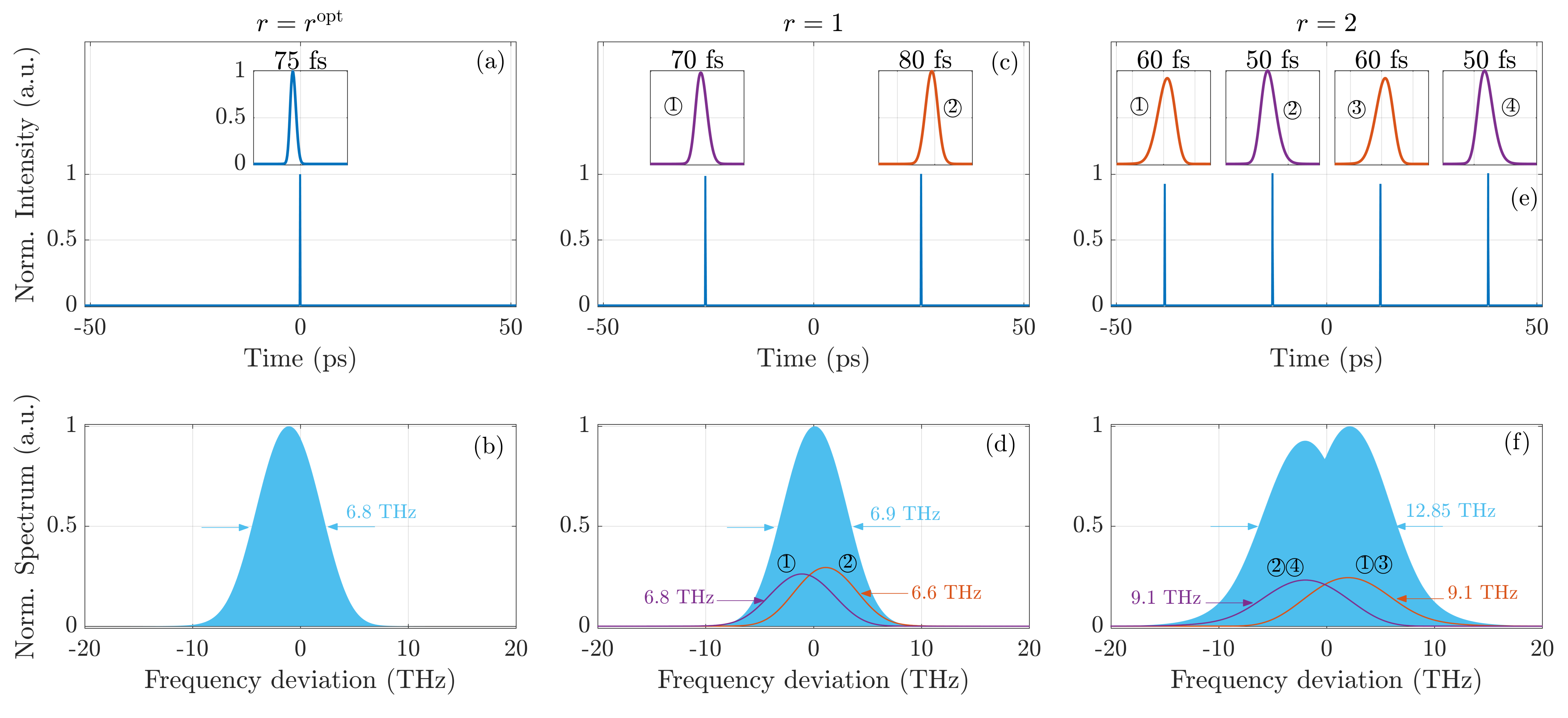}
    \caption{Pulse formation in the normal dispersion regime for various values of $r$. (top panels) Normalized intensity in the time domain along the full round-trip time ($\approx 100$~ps), showing the formation of femtosecond pulses detailed in the insets. (bottom panels) The corresponding normalized spectra. Curves in panels (d) and (f) are the spectra of single pulses, that correspond to the pulses in panels (c) and (d), respectively, thorough the circled numbers.}
    \label{fig:pulses532}
\end{figure*}

Figure~\ref{fig:pulses532} shows the intracavity pulses obtained for the pair of wavelengths 532-1064~nm, and for $r=r^{\mathrm{opt}},1,2$, where the signal wavelength lies in the normal dispersion regime. The electric field over the full round-trip is shown in the panels (a), (c) and (e), where over each pulse an inset is shown with the corresponding full-width at half-maximum (FWHM) expressed in femtoseconds. The panels (b), (d) and (f) show the spectra with the corresponding FWHM bandwidths in THz. As can be seen, the number of pulses obtained for each specific value of $r$ is consistent with Figure~\ref{fig:reqthreshold}. This is an important result demonstrating the control of the pulse repetition rate simply by varying the EOM frequency. It can also be seen in the top panels that as $r$ increases, the duration of the pulses decreases. This is consistent with the threshold condition where, for a fixed pump level, the time window in which there is transfer of photons from the pump to the signal becomes increasingly narrower. The curves in panels (d) and (f) correspond to the spectrum of a single pulse, labeled in panels (c) and (e) with circled numbers. The sum of the single spectra results in the full shaded spectrum.  For the case of $r=1$, one single pulse in linked with a single curve in panel (d), whilst for the case of $r=2$, the pair of pulses $\textcircled{1}$/$\textcircled{3}$ and $\textcircled{2}$/$\textcircled{4}$ share the same spectrum, shown in panel (f).

\begin{figure*}[htbp]
    \centering
    \includegraphics[width=1.0\textwidth]{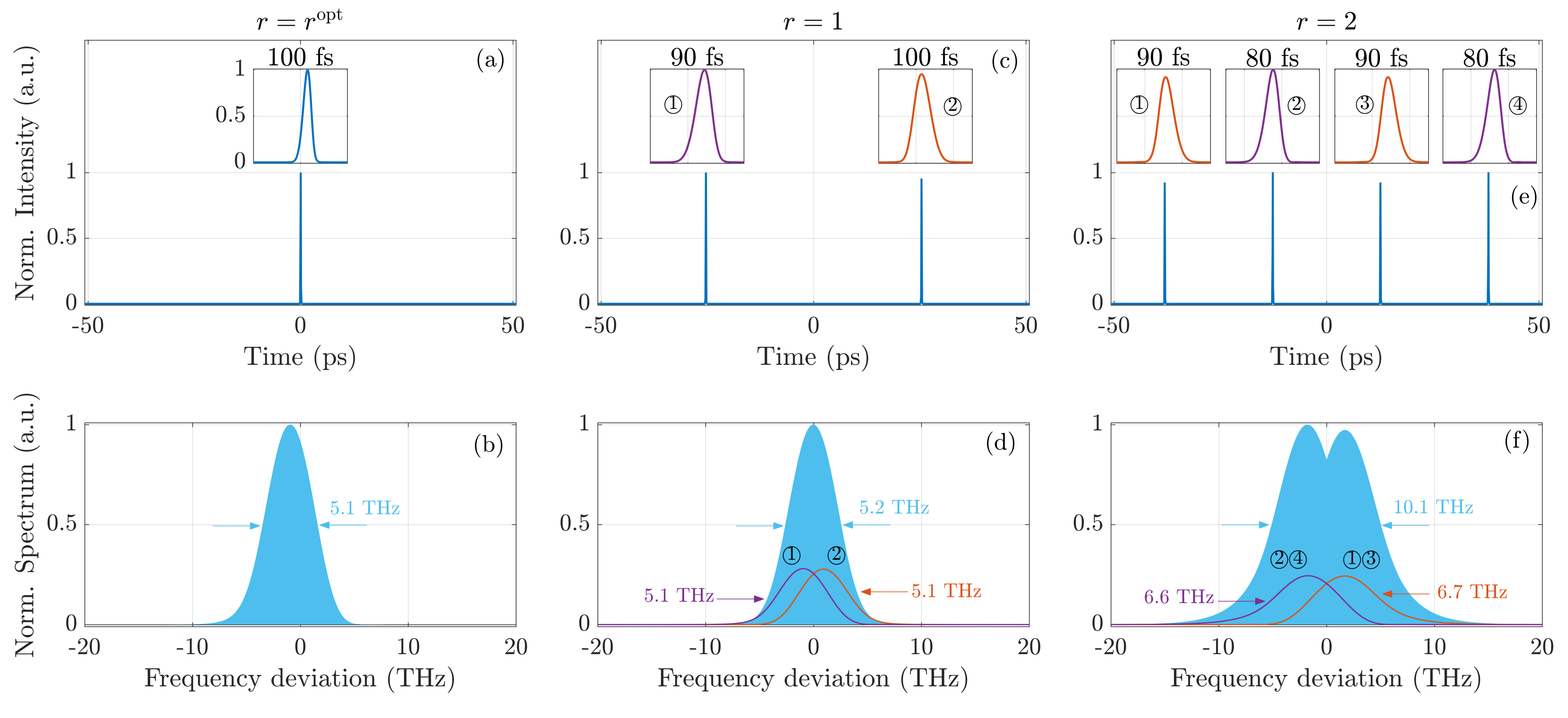}
    \caption{Pulse formation in the anomalous dispersion regime for various values of $r$. (top panels) Normalized intensity in the time domain along the full round-trip time ($\approx 100$~ps), showing the formation of femtosecond pulses detailed in the insets. (bottom panels) The corresponding normalized spectra. Curves in panels (d) and (f) are the spectra of single pulses, that correspond to the pulses in panels (c) and (d), respectively, thorough the circled numbers.}
    \label{fig:pulses1550}
\end{figure*}

Similarly, Figure~\ref{fig:pulses1550} shows the intracavity pulses obtained for the pair of wavelengths 1550-3100~nm, and for $r=r^{\mathrm{opt}},1,2$. In this case, the signal wavelength lies in the anomalous dispersion regime, in which the system shows a similar behaviour to that in the normal dispersion regime.

\begin{figure*}[htbp]
    \centering
    \includegraphics[width=1.0\textwidth]{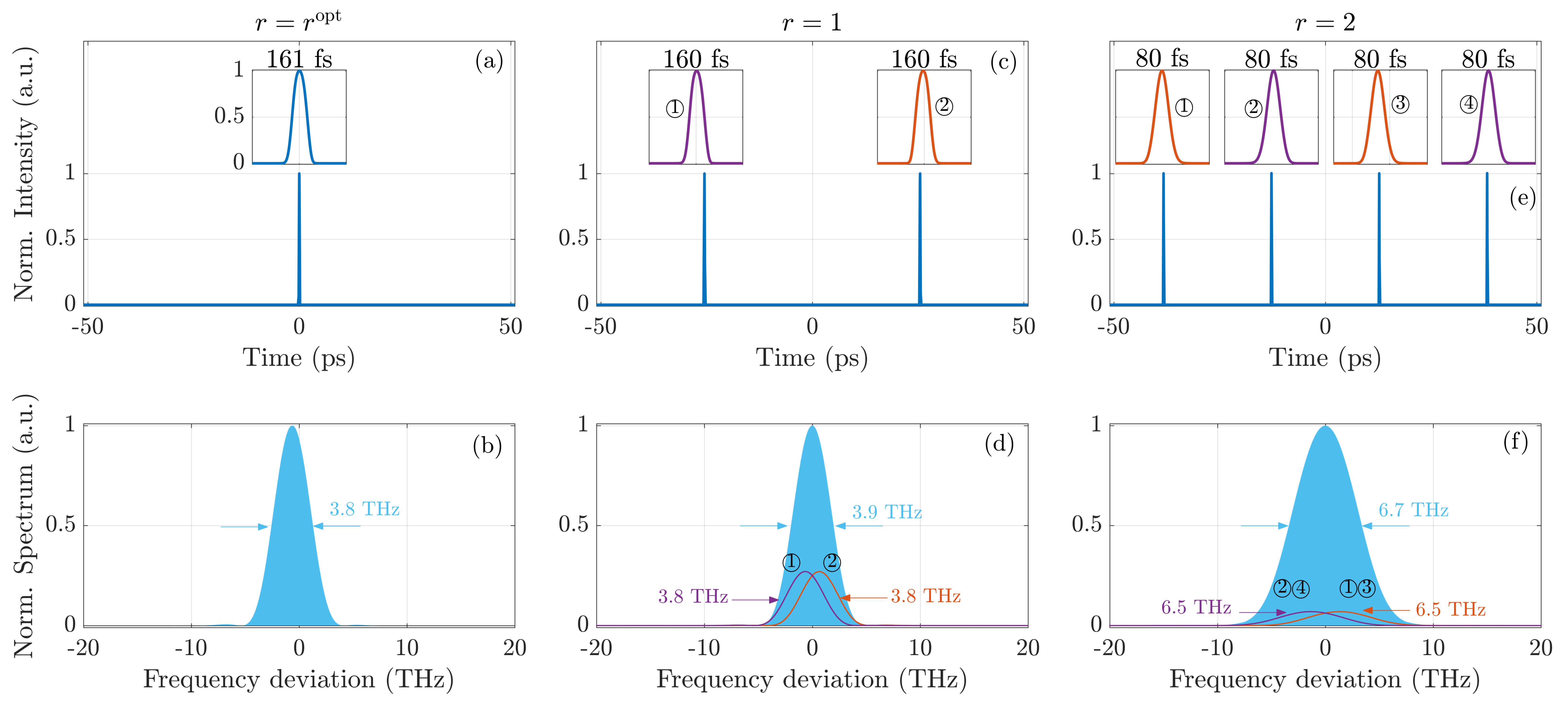}
    \caption{Pulse formation in the zero GVM regime for various values of $r$. (top panels) Normalized intensity in the time domain along the full round-trip time ($\approx 100$~ps), showing the formation of femtosecond pulses detailed in the insets. (bottom panels) The corresponding normalized spectra. Curves in panels (d) and (f) are the spectra of single pulses, that correspond to the pulses in panels (c) and (d), respectively, thorough the circled numbers.}
    \label{fig:pulses1351}
\end{figure*}

It can also be seen from the results obtained in Figures~\ref{fig:pulses532}~and~\ref{fig:pulses1550} that by varying $r$, a controllable repetition rate can be obtained in both the normal and anomalous dispersion regimes. However, it is evident that the pulses are not symmetric, and are also generated in pairs with different duration. Figure~\ref{fig:pulses1351} shows the result for the case in which $\dk^{\prime}= 0$, \edited{where the pulses obtained exhibit a symmetric shape and all are of equal duration for a given value} of $r$.

\begin{figure*}[htbp]
    \centering
    \includegraphics[width=1.0\textwidth]{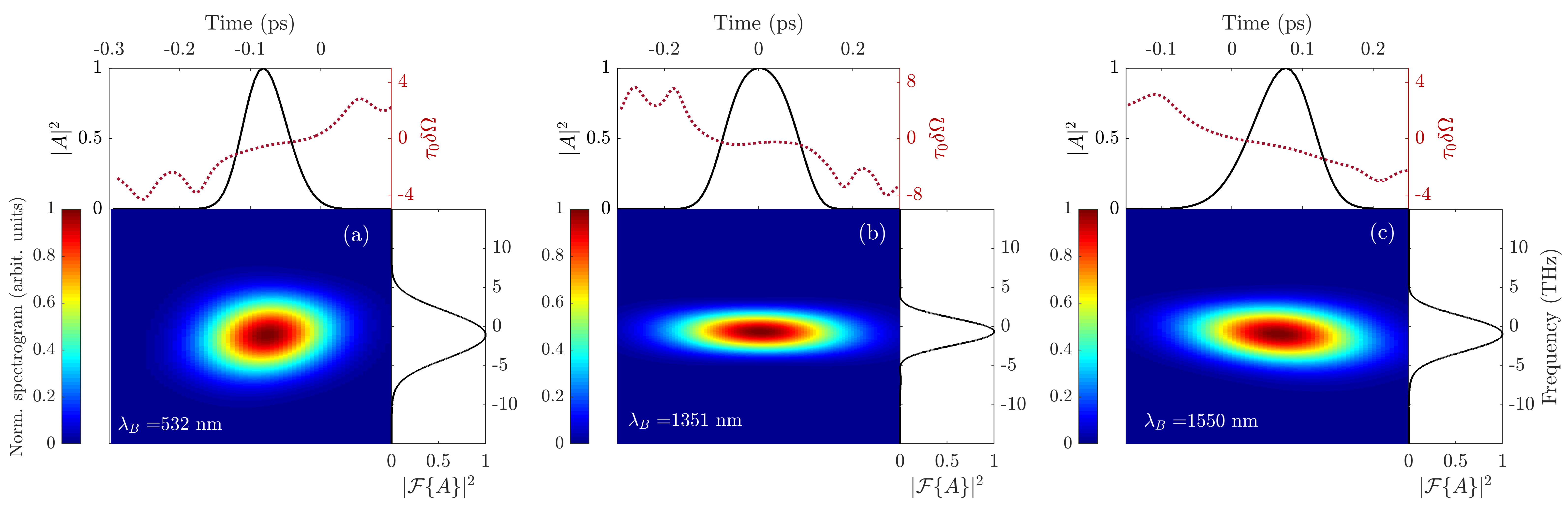}
    \caption{Spectrograms corresponding to the single pulses obtained when  $r=r^{\mathrm{opt}}$, for the relevant wavelengths. Each panel includes the temporal and spectral solution, together with the corresponding frequency chirp.}
    \label{fig:spect}
\end{figure*}

The asymmetry in the pulses for non-zero GVM ($\dk^{\prime}\neq 0$) owing to the nonlinear response, $\hat{I}(\Omega)$, induces a chirp, similar to Raman response in Kerr media~\cite{osborne1989raman,hook1989effects,schadt1987frequency}. 
Figure~\ref{fig:spect} shows the spectrogram of the single pulses obtained when $r=r^{\mathrm{opt}}$, for the three wavelengths of interest. In each panel, the spectrum (right), the intensity as a function of time (top, left axis) and the frequency chirp weighted by the pulse duration, $\tau_0$ (top, right axis) are also plotted. The frequency chirp is defined as $\delta\Omega=-d\phi_A(\tau)/d\tau$. Mathematically the spectrogram is defined as the Gabor transform~\cite{MELCHERT2019100275}
\begin{equation}
    \mathcal{S}(\tau^{\prime},\Omega) = \int_{-\infty}^{+\infty} A(\tau)w(\tau-\tau^{\prime})e^{-i\Omega\tau} \,d\tau,
\end{equation}
where $w(\tau-\tau^{\prime})$ is a variable-gate function. We used a Gaussian window as the variable-gate function with a FWHM of 70\% of the pulse width. Figure~\ref{fig:spect} shows the normalized $\abs{\mathcal{S}(\tau^{\prime},\Omega)}^2$ for a single pulse obtained from the simulations.
For example, if the pulse were an ideal unchirped pulse, the contour lines of $\mathcal{S}$ will describe non-rotated ellipses~\cite{kane1993characterization}. However, despite the fact that the spectrograms slightly deviates from perfect ellipses exhibiting a marginal chip, it shows that the generated pulses for zero GVM in Figure~\ref{fig:spect}(b) are nearly transformed limited.

\section{Conclusions}
\label{sec:concl}

In this work we have derived a mean-field equation (MFE) that describes the temporal and spectral evolution of the signal electric field in a degenerate cw-driven $\chi^{(2)}$ OPO containing an EOM and including intracavity dispersion compensation. The structure of the equation is consistent with the MFEs reported in previous works. Due to the complexity of this equation, the analytical solutions have not yet been obtained, and must be studied numerically. Based on the experimental feasibility of the proposed model, whether in bulk or integrated format, we have numerically solved the MFE and showed that ultrashort pulses ($<200$~fs) can be obtained from such a scheme and, in addition, the repetition rate can be controlled by varying the frequency of the EOM. In turn, the EOM controls the threshold condition of the OPO, enabling the pump field to transfer photons to the signal field in a very short time window, leading to ultrashort pulse formation.
We have also shown that the pulses can be obtained in both normal and anomalous dispersion regimes and that the value of GVM only influences the temporal and spectral asymmetries of the output pulses. We believe that the predictions of this new MFE will motivate the development and implementation of the proposed scheme for the realization of new sources of broadly tunable femtosecond pulses with variable repetition rates and in arbitrary wavelength regions using widely available cw lasers.

\begin{acknowledgments}
We gratefully acknowledge funding from the Ministerio de Ciencia e Innovación (MCIN) and the State Research Agency (AEI), Spain (Project Nutech PID2020-112700RB-I00); Project Ultrawave EUR2022-134051 funded by MCIN/AEI and by the “European Union NextGenerationEU/PRTR; Severo Ochoa Programme for Centres of Excellence in R\&D (CEX2019-000910-S);  Generalitat de Catalunya (CERCA); Fundación Cellex; Fundació Mir-Puig. S. Chaitanya Kumar acknowledges support of the Department of Atomic Energy, Government of India, under Project Identification No. RTI 4007. The authors would also like to express their sincere gratitude to Dr. Nicolas Linale for his insightful discussions about several topics this work includes. 
\end{acknowledgments}

\appendix*
\section{Numerical simulations}
Equation~\ref{eq:LLE_GDD_EOM} was solved using the the fourth-order Runge-Kutta method, solving the dispersive term and convolutions in the frequency domain. Our implementation is scripted in CUDA language and uses a graphics process unit (GPU) to speed up the computational times. Table~\ref{tab:Table1} shows the parameters used in the simulations performed in this work. The refractive index of MgO:PPLN nonlinear crystal at the respective wavelength is estimated from the relevant Sellemeier equations~\cite{gayer2008temperature}.
\begin{table*}[htbp]
\centering
\caption{\label{tab:Table1}Parameters used for our simulations.}
\vspace{5mm}
\begin{ruledtabular}
\begin{tabular}{rrl}
\hline
\textrm{\textbf{Parameter}} & \textrm{\textbf{Values}\footnote{The quantities with three values correspond to the pump wavelength at 532~nm (left), 1351~nm (center), and 1550~nm (right).}} & \textrm{\textbf{Units}}\\

Pump wavelength, $\lambda_B$            & 532/1351/1550     & nm                   \\
Pump beam waist, $w_{0p}$               & 55      & $\upmu$m                  \\
Degenerate wavelength, $\lambda_A$      & 1064/2702/3100    & nm                   \\
Cavity net reflectivity, $R$            & 0.98       & -                      \\
Cavity length, $\lcav$                  & 25      & mm                        \\
Round trip time, $\trt$                 & 102.55/101.98/101.55  & ps                 \\ 
$n_A$                                   & 2.15/2.11/2.09    & -                    \\
$n_B$                                   & 2.23/2.15/2.13    & -                    \\
$\dk^{\prime}$                          & 792/0/-60     & ps m$^{-1}$           \\
$k^{''}_A$                              & 0.235/-0.311/-0.561 & ps$^2$m$^{-1}$       \\
$k^{''}_B$                              & 0.719/0.151/0.0982 & ps$^2$m$^{-1}$       \\
$\deff$                                 & 14.77   & pm V$^{-1}$               \\
Crystal length, $\lcr$                  & 5       & mm                        \\
Grating period, $\Lambda$\footnote{We assumed the quasi phase-matching condition in our simulations: $\dk= 2k(\omega_0)-k(2\omega_0)-2\pi/\Lambda$}    & 6.95/35.9/35.01    & $\upmu$m            \\
Crystal temperature                     & 37.43/77.82/32.88     & $^{\circ}$C          \\ 
Time step ($\Delta \tau$)               & $\approx 6$    & fs \\
Number of point per round-trip          & $2^{14}$      & -                    \\
Round trips per simulation              & $20000$   & -                       \\ 
Execution time \footnote{We used for all simulations a GPU card NVIDIA, model GeForce GTX 1650. However, the execution time varies depending on the GPU used.}             & $\approx1000$    & round trips min$^{-1}$       \\ 
\end{tabular}
\end{ruledtabular}
\end{table*}

\nocite{*}

\bibliography{apssamp}

\end{document}